\documentclass[12pt]{article}
\usepackage[colorlinks=true,citecolor=blue,linkcolor=blue]{hyperref}
\usepackage{multirow}
\usepackage[left=2cm,right=2cm,top=2cm,bottom=2cm]{geometry}
\usepackage{cite}
\usepackage{psfrag}
\usepackage[demo]{graphicx}
\usepackage{graphicx}
\usepackage{graphics}
\usepackage{bigints}
\usepackage{epsfig}
\usepackage{caption,stackengine}
\usepackage{booktabs}
\usepackage{amsmath,amsfonts,amssymb}
\usepackage{pstcol,pst-fill,pst-grad}
\usepackage{pstricks,pst-fill,pst-grad}
\usepackage{euscript}
\usepackage{pstricks}
\usepackage{wrapfig}
\usepackage{subcaption}
\usepackage{slashed}
\usepackage[toc,page]{appendix}
\usepackage{here}

\begin{document}
	\title{\vspace{-3cm}
		\hfill\parbox{4cm}{\normalsize \emph{}}\\
		\vspace{1cm}
		{Muon pair production via $e^{+}e^{-}$ annihilation in the presence of a circularly polarized laser field}}
	\vspace{2cm}

	\author{M. Ouali,$^1$ M. Ouhammou,$^1$ S. Taj,$^1$ R. Benbrik,$^2$ and B. Manaut$^{1,}$\thanks{Corresponding author, E-mail: b.manaut@usms.ma} \\
		{\it {\small$^1$ Recherche Laboratory in Physics and Engineering Sciences,}}\\
		{\it {\small  Modern and Applied Physics Team, FPBM, USMS, Morocco.}}\\
		{\it {\small$^2$ High Energy Physics and Astrophysics Laboratory, FSSM, UCAM, Morocco.}}\\		
	}
	\maketitle \setcounter{page}{1}
\date{}
\begin{abstract}
In this paper, we have investigated the elementary particle reaction ${e}^{+} {e}^{-} \rightarrow{\mu}^{+} {\mu}^{-}$ that results from the electron-positron interaction, at the leading order, with an intense laser wave of circular polarization. We have derived, by analytical means, the laser-assisted differential cross section expression by using the scattering matrix approach. We have analyzed the energy and the number of exchanged photons dependence of muon pair production in electron-positron annihilation at different centre of mass energies including the $Z$-boson peak. For this reason, a wide range of high centre of mass energies relevant to future $e^{+}e^{-}$ collider were covered to study the cross section behavior. We have found that, for a given number of exchanged photons, laser field strength and frequency, the circularly polarized laser field decreases the total cross section by several orders of magnitudes.
\end{abstract}
\maketitle
\section{Introduction}
Since its introduction in 1960, studies in the field of laser-matter interaction usually deals with non relativistic atomic physics \cite{1}. Then, Some researchers have introduced laser-atom interaction in relativistic regime \cite{2,3}.
In recent years and due to the great technological advances \cite{4}, there exist powerful laser sources where the electrons and positrons can acquire high kinetic energies which lies far beyond the typical atomic energy scale. Therefore, such high energies can be exploited to induce heavy elementary particle reactions such as Higgs-strahlung production \cite{5,6}, charged Higgs pair production \cite{7}, neutral Higgs pair production \cite{8}, and heavy lepton-pair \cite{9,10} or hadron production in $e^{+}e^{-}$ colliders. In addition, lasers can be utilized to generate well-controlled particle interactions at microscopically small impact parameters, which can lead to high luminosity. Consequently, strong laser fields may provide alternative and complementary ways for high energy physics. In \cite{11}, we have studied the effect of the electromagnetic field with circular polarization on the $Z$-boson production, and it is found that the laser field decreases its cross section. In \cite{7,8}, we have studied the cross section of charged and neutral Higgs pair production at $e^{+}e^{-}$ collider  in the presence of a circularly polarized laser field.

It is well known that there exist two types of laser-matter interactions. The first type is called laser-induced interaction \cite{12} which can be induced by the presence of the laser field and can not occur without it. However, the second type which is named as laser-assisted process \cite{13,14,15,16,17}, occurs either in the presence or absence of the laser field. Therefore, muon pair production at $e^{+}e^{-}$ colliders is studied as a laser-assisted process where the electron and positron annihilate each other inside the electromagnetic field.

Future ${e}^{+} {e}^{-}$ colliders such as linear colliders (ILC ; CLIC) \cite{ILC-CLC} and circular colliders (FCC ; CEPC) \cite{FCC-CEPC} offer a rich physics program to test the standard model with a high precision measurement and search for evidences of new physics beyond the standard model. Electron-positron annihilation is a process studied with great attention in high energy physics, and it is widely used for experimental purposes. It allows the study of electroweak aspects, and it allows a whole series of studies on different aspects of QCD as well as the search for new particles. Muon-antimuon pair production from $e^{+}e^{-}$ annihilation is one of the most elementary processes in particle physics as it has proven fundamental for the understanding of other electron-positron interactions. The leading-order QED cross section of ${e}^{+} {e}^{-}\rightarrow {\mu}^{+} {\mu}^{-}$ is well established in the presence of both circularly \cite{circ} and linearly \cite{linear} polarized electromagnetic field. Moreover, it is found that the insertion of a circularly polarized laser field decreases the cross section while the linearly polarized laser field enhances it. However, the $Z$-boson Feynman diagram is essential as it has a great contribution, then it can not be neglected.
In this respect, we have investigated the process ${e}^{+} {e}^{-} \rightarrow{\mu}^{+} {\mu}^{-}$ at the lowest order, via the virtual photon ($\gamma$) and $Z$-boson exchange, in the presence a circularly polarized electromagnetic field. To the best of our knowledge, the process ${e}^{+} {e}^{-} \rightarrow (\gamma, Z)\rightarrow{\mu}^{+} {\mu}^{-}$ in the presence of an external field has not been considered before. The aim of this paper is to illustrate the effect of the laser field on the process of muon pair production at $e^{+}e^{-}$ collider from the theoretical measurement of the cross section as a function of the centre of mass energy ($\sqrt{s}$) and the laser field's parameters such as laser's electric field amplitude, its frequency and the number of exchanged photons. 
 
The remained of this paper is organized as follows: The next section is devoted to the description of the electromagnetic field, the wave functions of the interacting particles and the theoretical calculation of the differential cross section. Section 3 deals with the analysis of the obtained data and results. A short conclusion is given in section 4. We mention that, throughout this paper,  we have used natural units such that $\hbar=c=1$. The choice made for Livi-Civita tensor is that $\epsilon^{0123}=1$ and the metric $g^{\mu\nu}$ is taken such that $g^{\mu\nu}=(1,-1,-1,-1)$.

\section{Outline of the theory}\label{sec:theory} 
\subsection{Description of the laser field and wave functions}
\begin{figure}[H]
  \centering
      \includegraphics[scale=0.33]{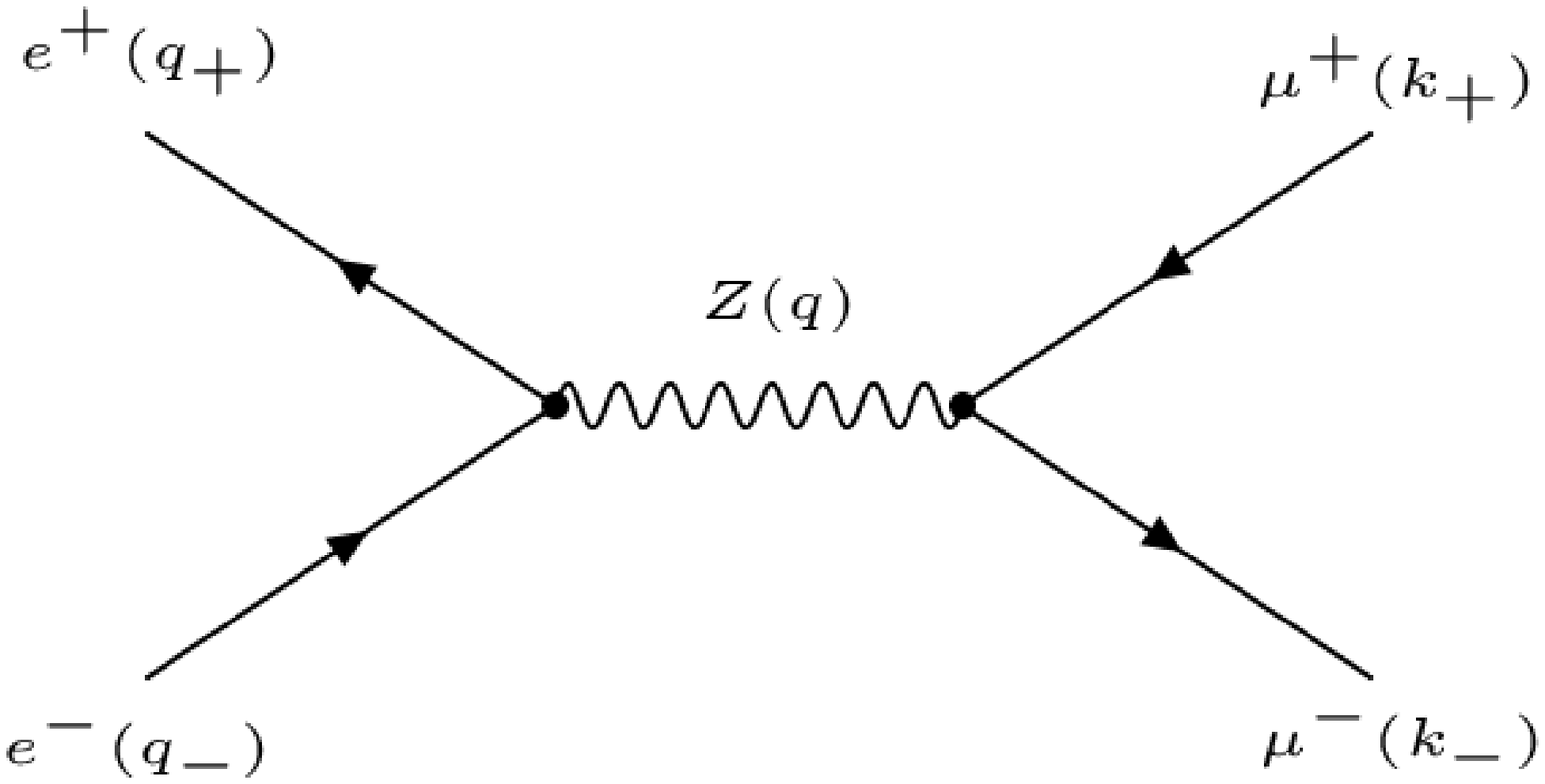}\hspace*{0.4cm}
      \includegraphics[scale=0.33]{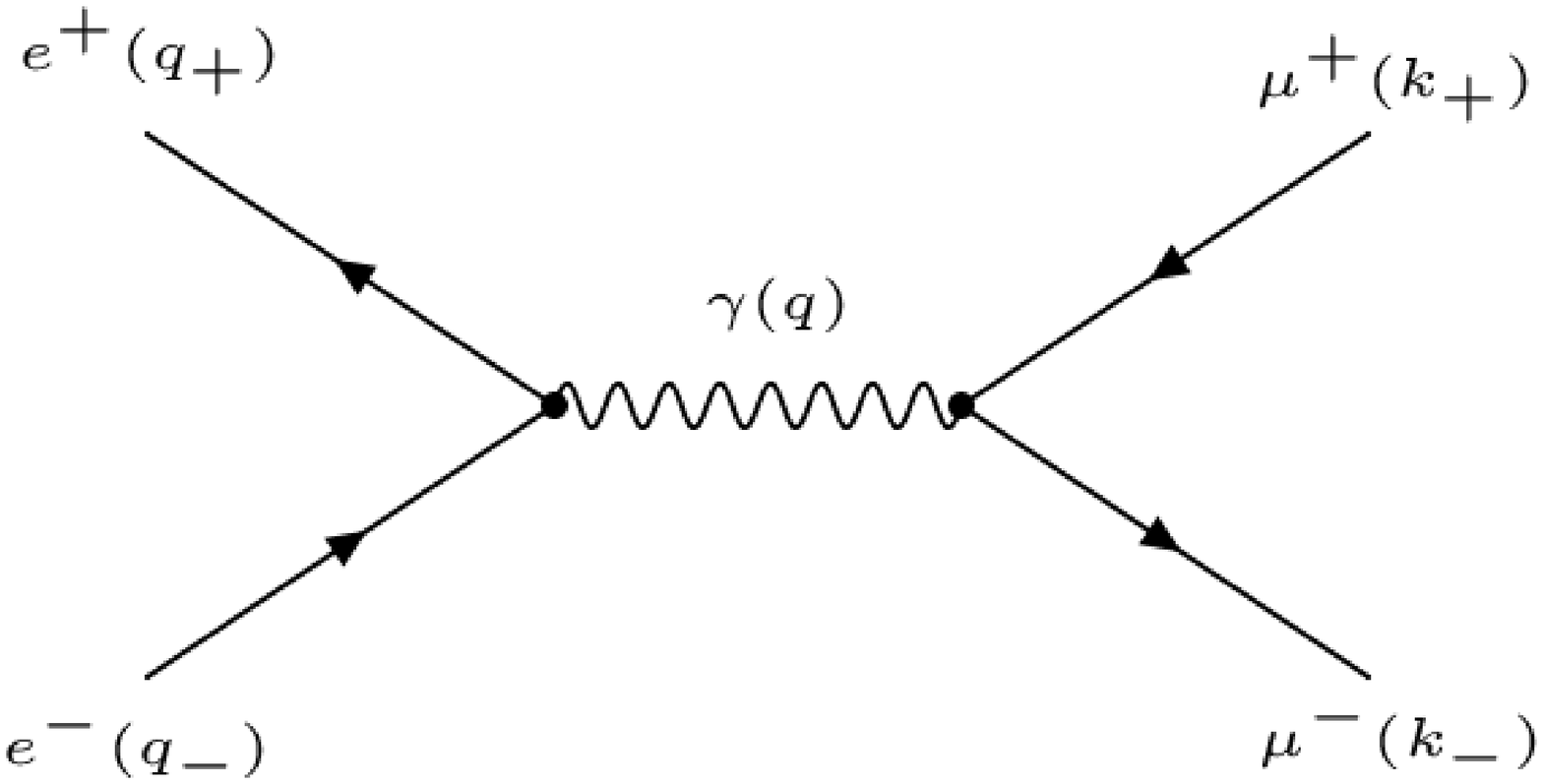}\par\vspace*{0.5cm}
        \caption{Leading-order Feynman diagrams for the electron-positron annihilation process $e^{-}e^{+}\rightarrow \mu^{-}\mu^{+}$.}
        \label{fig1}
\end{figure}

\begin{figure}[t]
  \centering
      \includegraphics[scale=1.2]{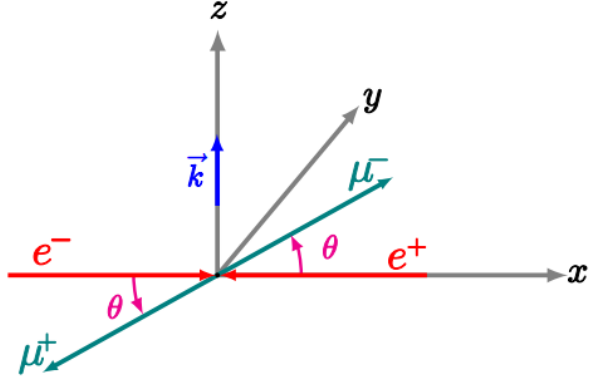}
  \caption{Geometry of the electron-positron annihilation $e^{+}e^{-}\rightarrow \mu^{+}\mu^{-}$ collision in the centre of mass frame.}
  \label{fig2}
\end{figure}

The process which acts as a source of muon pair production at the electron-positron colliders is denoted as: 
\begin{equation}
e^{-}(q_{-}, s_{-})+e^{+}(q_{+}, s_{+})\rightarrow \mu^{-}(k_{-},s_{-})+\mu^{+}(k_{+},s_{+}),
\label{1}
\end{equation}
where $q_{-}$ and $q_{+}$ are the effective momenta of the electron and positron. $k_{-}$ and $k_{+}$ are successively the free momenta of the muon and the antimuon.
The conversion of an electron-positron pair into a muon pair can occur with either a virtual photon, a $Z$ boson or the Higgs boson, in the intermediate state. All these processes are in principle indistinguishable, since they yield the same final state. Due to the fact that the contribution from the Higgs particle is totally negligible, we will focus on the $\gamma$ and $Z$ exchange processes, which are described by figure \ref{fig1}.
Throughout this paper, we have treated the produced muons as free particles while the electron and positron are embedded in a monochromatic electromagnetic field with circular polarization. Moreover, the only way to modify the free states of the electron and positron, in CEPC collider for example, is to position the electromagnetic beam as perpendicular to the colliding beam. Therefore, the collision is considered as taking place in the (xoy) while the laser field is propagating along the $z$-axis as illustrated in figure \ref{fig2}.
The classical four-potential of the laser field is given by:
\begin{equation}
A^{\mu}(\phi)=a_{1}^{\mu}\cos\phi+a_{2}^{\mu}\sin\phi,
\label{2}
\end{equation}
where $\phi=(k.x)$ is the phase of the electromagnetic field. $k^{\mu}=\omega(1,0,0,1)$ is the wave four vector, and $a_{1,2}^{\mu}$ are four-vectors chosen as $a_{1}^{\mu}=|a|(0,1,0,0)$ and $a_{2}^{\mu}=|a|(0,0,1,0)$, with $a$ denoting the amplitude of the vector potential. These quantities satisfy the following conditions: $a_{1}^{2}=a_{2}^{2}=a^{2}=-|\mathbf{a}|^{2}=-\big(\varepsilon_{0}/\omega \big)^{2}$ where $\varepsilon_{0}$ is the amplitude of the external field. From the Lorentz gauge condition $\partial_{\mu}A^{\mu}=0$, we deduce that $ a_{1}.k=0$ and $\, a_{2}.k=0$. Both electron and positron are described by the Dirac-Volkov states \cite{Volkov} such that:
\begin{equation}
\psi_{p_{\pm},s_{\pm}}(x)= \Big[1\pm\dfrac{e \slashed k \slashed A}{2(k.p_{\pm})}\Big] \frac{u(p_{\pm},s_{\pm})}{\sqrt{2Q_{\pm}V}} e^{iS(q_{\pm},s_{\pm})},
\label{3}
\end{equation}
with:
\begin{equation}
S(q_{\pm},s_{\pm})=\pm q_{\pm}x +\frac{e(a_{1}.p_{\pm})}{k.p_{\pm}}\sin\phi - \frac{e(a_{2}.p_{\pm})}{k.p_{\pm}}\cos\phi.
\label{4}
\end{equation}
In equation (\ref{3}), $u_{(p_{\pm}, s_{\pm})}$ are the Dirac spionrs, $s_{\pm}$ denote the particles spin, $x$ is the space-time coordinate of the incident particles, $Q_{\pm}$ are their acquired effective energies, $p_{\pm}=(E_{\pm},\mp \,p_{\pm},0,0)$ are the free four-momenta of the electron and positron outside the electromagnetic field such that:
\begin{equation}
q_{\pm}=p_{\pm}+\dfrac{e^{2}a^{2}}{2(k.p_{\pm})}k.
\label{5}
\end{equation}
The corresponding effective mass reads $m_{e}^{*}=\sqrt{(q_{i}^{2})}=\big(m_{e}^{2}+e^{2}a^{2}\big)^{\frac{1}{2}}$, with $e$ is the electron charge and $m_{e}$ denotes its mass outside the laser field. Like free state, the Dirac-Volkov states in equation (\ref{3}) are normalized to $\delta$-function. The produced muon-antimuon are considered as free. Therefore, they are described by free states \cite{Greiner} such that:
\begin{equation}
\psi_{k_{\pm},s_{\pm}}(y)= \frac{u(k_{\pm},s_{\pm})}{\sqrt{2E_{\mu^{\pm}}V}} e^{\pm ik_{\pm}y},
 \label{6}
\end{equation}
with $y$ denotes the space-time coordinate of the muon and the antimuon, $k_{-}(E_{\mu^{-}},|k_{-}|\cos \theta,|k_{-}|\sin \theta,0)$ and $k_{+}(E_{\mu^{+}},-|k_{+}|\cos \theta,-|k_{+}|\sin \theta,0)$ are their free momentum, and $E_{\mu^{-}}$ and $E_{\mu^{+}}$ are their corresponding energies.
\subsection{Transition amplitude and cross section}
\begin{figure}[H]
  \centering
      \includegraphics[scale=0.5]{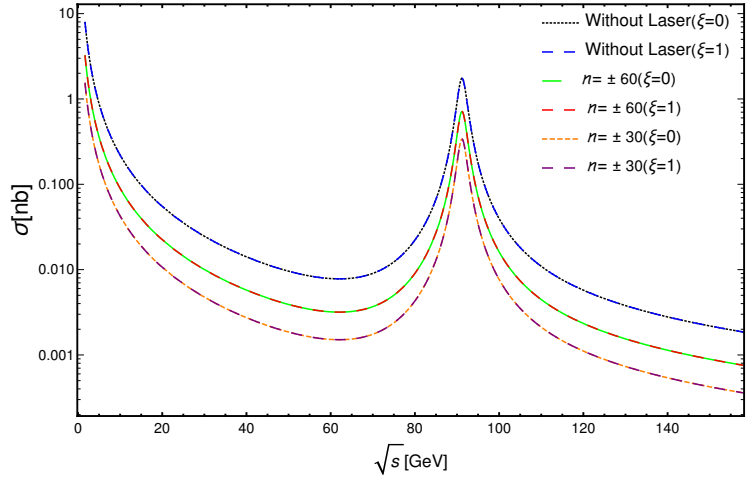}
  \caption{Dependence of the total cross section, for two gauges, on the centre of mass energy for different number of exchanged photons by choosing the laser field strength as $\varepsilon_{0}=10^{6}\,V.cm^{-1}$ and its frequency as $\omega=1.17\,eV$.}
  \label{fig3}
\end{figure}

According to the Feynman rules and by using the following coupling of $Z$ and $\gamma$ bosons to charged fermions:
 \begin{equation}
     \begin{cases}
    \gamma \mu^- \mu^+ = ie\gamma^{\mu} \quad \quad  ; \quad \quad  Z \mu^- \mu^+     =\frac{i\,g}{4\cos(\theta_W)}\gamma^{\mu}(g_{v}-g_{a}\gamma^{5}) \\ 
    \gamma  e^- e^+ = ie\gamma^{\mu} \quad \quad  ; \quad \quad  Z e^- e^+     =\frac{i\,g}{4\cos(\theta_W)}\gamma^{\mu}(g_{v}-g_{a}\gamma^{5}),
    \end{cases}
     \label{7}
 \end{equation}
the laser-assisted scattering matrix element \cite{Greiner} for muon pair production can be expressed as follows:
\small
\begin{eqnarray}
S_{fi}({e}^{+}{e}^{-}\rightarrow \mu^{+}\mu^{-})&=&\nonumber  \int_{}^{} d^4x \int d^4y \Bigg\lbrace\overline{\psi}_{p_{+},s_{+}}(x)\dfrac{i\,g}{4 \cos\theta_{w}}\gamma^{\mu}(g_{v}-g_{a}\gamma^{5})\psi_{p_{-},s_{-}}(x)D_{\mu\nu}(x-y)\\ &\times &\overline{\psi}_{k_{-},s_{-}}(y)\dfrac{i\,g}{4 \cos\theta_{w}} \gamma^{\nu}(g_{v}-g_{a}\gamma^{5}) \psi_{k_{+},s_{+}}(y)+\bar{\psi}_{p_{+},s_{+}}(x) (ie\gamma^{\mu}) \psi_{p_{-},s_{-}}(x)\nonumber \\ &\times & G_{\mu\nu}(x-y)  \bar{\psi}_{k_{-},s_{-}}(y) (ie\gamma^{\nu}) \psi_{k_{+},s_{+}}(y) \Bigg\rbrace.
\label{8}
\end{eqnarray}
\normalsize
$ D_{\mu \nu}(x-y) $ is the $Z$-boson propagator, and $G_{\mu \nu}(x-y)$ is the photon propagator \cite{Greiner}. Their expressions are given by:
\begin{equation}
D_{\mu\nu}(x-y)=\bigintsss \dfrac{d^{4}q}{(2\pi)^4} \frac{e^{-iq(x-y)}}{q^{2}-M_{Z}^{2}}\Bigg(-ig_{\mu\nu}+i(1-\xi)\dfrac{q_{\mu}q_{\nu}}{M_{Z}^{2}}\Bigg),
\label{9}
\end{equation}
\begin{equation}
G_{\mu\nu}(x-y)=\bigintsss \dfrac{d^{4}q}{(2\pi)^4} \frac{e^{-iq(x-y)}}{q^{2}}\Bigg(-ig_{\mu\nu}+i(1-\xi)\dfrac{q_{\mu}q_{\nu}}{q^{2}}\Bigg),
\label{10}
\end{equation}
where $q$ denotes their four-momentum. In equation (\ref{7}), $g_v=-1+4\sin^{2}\theta_{W} $ and $g_a=-1$ are the vector and axial vector coupling constants, $\theta_W$ is the Weinberg angle, {$ g $} is the electroweak coupling constant such that $g^{2}=e^{2}/\sin^{2}\theta_{W}=8G_{F}M_{Z}^{2}\cos^{2}_{\theta_{W}}/\sqrt{2}$. 
After substituting the equations (\ref{3}), (\ref{6}), (\ref{9}) and (\ref{10}) into the equation (\ref{8}), the space-time integration can be performed by the standard method of Fourier series expansion, using the generating function of the Bessel functions. The latter can be expressed via ordinary Bessel functions by using the Auger transformation \cite{Auger} such that:
\begin{equation}
e^{iz\sin\phi}=\sum_{n=-\infty}^{n=+\infty}J_{n}(z)e^{in\phi}.
\label{11}
\end{equation}
Therefore, the scattering matrix element becomes as follows:
\begin{eqnarray}
S_{fi}^{n}({e}^{+}{e}^{-}\rightarrow \mu^{+}\mu^{-})&=&\dfrac{(2\pi)^{4}\delta^{4}(k_{-}+k_{+}-q_{-}-q_{+}-nk)}{4V^{2}\sqrt{Q_{-}Q_{+}E_{\mu^{+}}E_{\mu^{-}}}} \big(  A_{\gamma}^{n} + A_{Z}^{n} \big).
\label{12}  
\end{eqnarray}
The integer number $n$ in equation (\ref{12}) counts the laser photons that are emitted (if $n > 0$) or absorbed (if $n < 0$) by the electron and positron. The $\delta^{4}(k_{-}+k_{+}-q_{-}-q_{+}-nk)$ guarantees the energy-momentum conservation.
It is obvious that the total scattering amplitude consists of two parts $A_{\gamma}^{n}$ and $A_{Z}^{n}$. The former comes from the contribution of the free-photon exchange while the latter stands for the $Z$-boson contribution. $A_{\gamma}^{n}$ can be expressed in terms of ordinary Bessel functions as follows:
\small
\begin{eqnarray}
A_{\gamma}^{n}&=& \frac{e^{2}}{(q_{-}+q_{+}+nk)^{2}}  \Bigg\lbrace\bar{u}(p_{+},s_{+})\Bigg[ \chi_{0}^{\mu}\,J_{n}(z)e^{-in\phi _{0}}(z) +\frac{1}{2} \,\, \chi_{1}^{\mu}\Big(J_{n+1}(z)e^{-i(n+1)\phi _{0}} \nonumber \\ &+ & \nonumber J_{n-1}(z)e^{-i(n-1)\phi _{0}}\Big) + \frac{1}{2\, i}\,\chi_{2}^{\mu} \Big(J_{n+1}(z)e^{-i(n+1)\phi _{0}}-J_{n-1}(z)e^{-i(n-1)\phi _{0}}\Big)\Bigg] u(p_{-},s_{-})\\ &\times &\Big(-g_{\mu\nu}+(1-\xi)\dfrac{(q_{-}+q_{+}+nk)_{\mu}(q_{-}+q_{+}+nk)_{\nu}}{(q_{-}+q_{+}+nk)^{2}}\Big)\Bigg(\bar{u}(k_{-},s_{-})\gamma^{\nu}u(k_{+},s_{+})\Bigg) \Bigg\rbrace,
\label{13}
\normalsize
\end{eqnarray}
where the quantities $\chi_{0}^{\mu}$, $\chi_{1}^{\mu}$ and $\chi_{2}^{\mu}$ can be expressed as follows:
\begin{equation}
\begin{cases}\chi_{0}^{\mu}=\gamma^{\mu}+2c_{p_{-}}c_{p_{+}}a^{2}k^{\mu}\slashed k
   &\\\chi_{1}^{\mu}=c_{p_{-}}\gamma^{\mu}\slashed k\slashed a_{1}-c_{p_{+}}\slashed a_{1}\slashed k \gamma^{\mu}   
   &\\\chi_{2}^{\mu}=c_{p_{-}}\gamma^{\mu}\slashed k\slashed a_{2}-c_{p_{+}}\slashed a_{2}\slashed k \gamma^{\mu}
\end{cases}
\label{14}
\end{equation}
Analogously, we find that $A_{Z}^{n}$ can be expressed as follows:
\small
\begin{eqnarray}
A_{Z}^{n}&=& \dfrac{g^{2}}{16 \cos^{2}\theta_{w}}  \frac{1}{(q_{-}+q_{+}+nk)^{2}-M_{Z}^{2}}  \Bigg\lbrace\bar{u}(p_{+},s_{+})  \Bigg[ \lambda_{0}^{\mu}\,J_{n}(z)e^{-in\phi _{0}}(z) + \frac{1}{2} \,\, \lambda_{1}^{\mu}\Big(J_{n+1}(z)\\ &\times &\nonumber e^{-i(n+1)\phi _{0}} + J_{n-1}(z)e^{-i(n-1)\phi _{0}}\Big)  + \frac{1}{2\, i}\, \lambda_{2}^{\mu}\Big(J_{n+1}(z)e^{-i(n+1)\phi _{0}}-J_{n-1}(z)e^{-i(n-1)\phi _{0}}\Big)\Bigg] \\ &\times&\nonumber u(p_{-},s_{-}) \Big(-g_{\mu\nu}+(1-\xi)\dfrac{(q_{-}+q_{+}+nk)_{\mu}(q_{-}+q_{+}+nk)_{\nu}}{M_{Z}^{2}}\Big) \Bigg(\overline{u}(k_{-},s_{-})\gamma^{\nu}(g_{v}-g_{a}\gamma^{5})u(k_{+},s_{+})\Bigg)  \Bigg\rbrace,
\label{15}
\end{eqnarray}
\normalsize
where the expressions of $\lambda_{0}^{\mu}$, $\lambda_{1}^{\mu}$ and $\lambda_{2}^{\mu}$ are given by:
\begin{equation}
\begin{cases}
\lambda_{0}^{\mu}=\gamma^{\mu}(g_{v}-g_{a}\gamma^{5})+2c_{p_{-}}c_{p_{+}}a^{2}k^{\mu}\slashed k(g_{v}-g_{a}\gamma^{5})   &\\ \lambda _{1}^{\mu}=c_{p_{-}}\gamma^{\mu}(g_{v}-g_{a}\gamma^{5})\slashed k\slashed a_{1}-c_{p_{+}}\slashed a_{1}\slashed k \gamma^{\mu}(g_{v}-g_{a}\gamma^{5})   &\\ \lambda_{2}^{\mu}=c_{p_{-}}\gamma^{\mu}(g_{v}-g_{a}\gamma^{5})\slashed k\slashed a_{2}-c_{p_{+}}\slashed a_{2}\slashed k \gamma^{\mu}(g_{v}-g_{a}\gamma^{5}),
\end{cases}
\label{16}
\end{equation}
with $c_{p_{\pm}}=e/2(kp_{\pm})$. The argument of the Bessel function, $z$, and its phase $\phi_{0}$ are given by: $ z=\sqrt{\eta_{1}^{2}+\eta_{2}^{2}}$ and $\phi_{0}= \arctan(\eta_{2}/\eta_{1})$, with:
\begin{center}
$\eta_{1}=\dfrac{e(a_{1}.p_{-})}{(k.p_{-})}-\dfrac{e(a_{1}.p_{+})}{(k.p_{+})}$ \qquad ; \qquad $\eta_{2}=\dfrac{e(a_{2}.p_{-})}{(k.p_{-})}-\dfrac{e(a_{2}.p_{+})}{(k.p_{+})}$.\\
\end{center}
In the centre-of-mass frame, the differential cross section can be calculated by squaring the scattering matrix element given by equation (\ref{12}) and dividing the result by $V T$ to obtain the transition probability per volume, by $J$ , and by the particle density $\rho = V^{-1}$. Finally we have to integrate over the final states for a fixed emission angle $d\Omega=\sin \theta d\theta d\phi$. The differential cross section can be expressed as follows:
\begin{equation}
d\sigma_{n}=\dfrac{|S_{fi}^{n}|^{2}}{VT}\frac{1}{|J_{inc}|}\frac{1}{\varrho}V\int_{}\dfrac{d^{3}k_{-}}{(2\pi)^3}V\int_{}\dfrac{d^{3}k_{+}}{(2\pi)^3},
\label{17}
\end{equation}
where $|J_{inc}|=(\sqrt{(q_{-}q_{+})^{2}-m_{e}^{*^{4}}}/{Q_{-}Q_{+}V})$ denotes the current of incident particles in the centre of mass frame.
After simplifications and by averaging over the polarizations of the incoming particles, and summing over the final ones, we get:
\small
\begin{eqnarray}
d\sigma_{n}({e}^{+}{e}^{-}\rightarrow \mu^{+}\mu^{-})&=&\dfrac{1}{16\sqrt{(q_{-}q_{+})^2-m_{e}^{*^{4}}}}   \big|\overline{A_{\gamma}^{n} + A_{Z}^{n}} \big|^{2} \int_{}\dfrac{|\mathbf{k}_{-}|^{2}d|\mathbf{k}_{-}|d\Omega}{(2\pi)^2E_{\mu^{-}}}\int_{}\dfrac{d^{3}k_{+}}{ E_{\mu^{+}}} \nonumber \\  &\times & \delta^{4}(k_{-}+k_{+}-q_{-}-q_{+}-nk).
\label{18}
\end{eqnarray}
\normalsize
The spin-averaged square of the matrix elements $A_{\gamma}^{n}$ and $A_{Z}^{n}$ can be rewritten as traces of Dirac matrices in the usual manner by using the formulas $\sum_{s_{-}}^{}u(p_{-},s_{-})\bar{u}(p_{-},s_{-})=(\slashed p_{-}+m_{e})$ and $\sum_{s_{+}}^{}u(p_{+},s_{+})\bar{u}(p_{+},s_{+})=(\slashed p_{+}-m_{e})$, with $\slashed p_{\pm}=\gamma_{\mu}p_{\pm}^{\mu}$. Thus, we find that:
\tiny
\begin{eqnarray}
 \big|\overline{A_{\gamma}^{n} + A_{Z}^{n}} \big|^{2}&=&\nonumber\dfrac{1}{4}\sum_{n=-\infty}^{+\infty}\sum_{s_{-}s_{+}}\big| A_{\gamma}^{n} + A_{Z}^{n} \big|^{2}=\frac{1}{4}\sum_{n=-\infty}^{+\infty}\Bigg\lbrace\dfrac{e^{4}}{(q_{-}+q_{+}+nk)^{4}}\Big(-g_{\mu\nu}+(1-\xi)\dfrac{(q_{-}+q_{+}+nk)_{\mu}(q_{-}+q_{+}+nk)_{\nu}}{(q_{-}+q_{+}+nk)^{2}}\Big)\\ &\times & \nonumber\Big(-g_{\rho\beta}+(1-\xi)\dfrac{(q_{-}+q_{+}+nk)_{\rho}(q_{-}+q_{+}+nk)_{\beta}}{(q_{-}+q_{+}+nk)^{2}}\Big) Tr\Bigg[(\slashed p_{-}-m_{e})\Big[ \chi_{0}^{\mu}\,J_{n}(z)e^{-in\phi _{0}}(z)\\ \nonumber &+ &\chi_{1}^{\mu}\,\,\frac{1}{2}\Big(J_{n+1}(z)e^{-i(n+1)\phi _{0}} + J_{n-1}(z)e^{-i(n-1)\phi _{0}}\Big) +\chi_{2}^{\mu}\,\frac{1}{2\, i}\Big(J_{n+1}(z)e^{-i(n+1)\phi _{0}}- J_{n-1}(z)e^{-i(n-1)\phi _{0}}\Big)\Big]\\ &\times & \nonumber(\slashed p_{+}+m_{e}) \Big[ \chi_{0}^{\rho}\,J^{*}_{n}(z)e^{+in\phi _{0}}(z)+\chi_{1}^{\rho}\,\,\frac{1}{2}\Big(J^{*}_{n+1}(z)e^{+i(n+1)\phi _{0}} + J^{*}_{n-1}(z)e^{+i(n-1)\phi _{0}}\Big) \\ &- & \nonumber\chi_{2}^{\rho}\,\frac{1}{2\, i}\Big(J^{*}_{n+1}(z)e^{+i(n+1)\phi _{0}}-J^{*}_{n-1}(z)e^{+i(n-1)\phi _{0}}\Big)\Big]\Bigg] Tr\Bigg[ (\slashed k_{-}-m_{\mu})\gamma^{\nu} (\slashed k_{+}+m_{\mu})\gamma^{\beta} \Bigg]
+\left(\dfrac{g^{2}}{16 \cos^{2}\theta_{w}}\right)^{2}\\ &\times & \nonumber \left(\frac{1}{(q_{-}+q_{+}+nk)^{2}-M_{Z}^{2}}\right)^{2} \Big(-g_{\mu\nu}+(1-\xi)\dfrac{(q_{-}+q_{+}+nk)_{\mu}(q_{-}+q_{+}+nk)_{\nu}}{M_{Z}^{2}}\Big)\\ &\times & \nonumber \Big(-g_{\rho\beta}+(1-\xi)\dfrac{(q_{-}+q_{+}+nk)_{\rho}(q_{-}+q_{+}+nk)_{\beta}}{M_{Z}^{2}}\Big)  Tr\Bigg[(\slashed p_{-}-m_{e})\Big[ \lambda_{0}^{\mu}\,J_{n}(z)e^{-in\phi _{0}}(z)\\ &+ & \nonumber  \lambda_{1}^{\mu}\,\,\frac{1}{2}\Big(J_{n+1}(z)e^{-i(n+1)\phi _{0}} + J_{n-1}(z)e^{-i(n-1)\phi _{0}}\Big)+\lambda_{2}^{\mu}\,\frac{1}{2\, i}\Big(J_{n+1}(z)e^{-i(n+1)\phi _{0}}-J_{n-1}(z)e^{-i(n-1)\phi _{0}}\Big)\Big]\\ &\times & \nonumber  (\slashed p_{+}+m_{e}) \Big[ \lambda_{0}^{\rho}\,J^{*}_{n}(z)e^{+in\phi _{0}}(z)+ \lambda_{1}^{\rho}\,\,\frac{1}{2}\Big(J^{*}_{n+1}(z)e^{+i(n+1)\phi _{0}} + J^{*}_{n-1}(z)e^{+i(n-1)\phi _{0}}\Big) \\ &- & \nonumber \lambda_{2}^{\rho}\,\frac{1}{2\, i}\Big(J^{*}_{n+1}(z)e^{+i(n+1)\phi _{0}}-J^{*}_{n-1}(z)e^{+i(n-1)\phi _{0}}\Big)\Big]\Bigg] Tr\Bigg[ (\slashed k_{-}-m_{\mu})\gamma^{\nu}(g_{v}-g_{a}\gamma^{5}) (\slashed k_{+}+m_{\mu})\gamma^{\beta}(g_{v}-g_{a}\gamma^{5}) \Bigg]
\\ &+ & \nonumber  \dfrac{e^{2}}{(q_{-}+q_{+}+nk)^{2}}\left(\dfrac{g^{2}}{16 \cos^{2}\theta_{w}}\right)\frac{1}{(q_{-}+q_{+}+nk)^{2}-M_{Z}^{2}} \Big(-g_{\mu\nu}+(1-\xi)\dfrac{(q_{-}+q_{+}+nk)_{\mu}(q_{-}+q_{+}+nk)_{\nu}}{(q_{-}+q_{+}+nk)^{2}}\Big)\\ \nonumber &\times & \Big(-g_{\rho\beta}+(1-\xi)\dfrac{(q_{-}+q_{+}+nk)_{\rho}(q_{-}+q_{+}+nk)_{\beta}}{M_{Z}^{2}}\Big)  Tr\Bigg[(\slashed p_{-}-m_{e})\Big[ \chi_{0}^{\mu}\,J_{n}(z)e^{-in\phi _{0}}(z) \\ \nonumber &+ & \chi_{1}^{\mu}\,\,\frac{1}{2}\Big(J_{n+1}(z)e^{-i(n+1)\phi _{0}}+ J_{n-1}(z)e^{-i(n-1)\phi _{0}}\Big)+ \chi_{2}^{\mu}\,\frac{1}{2\, i}\Big(J_{n+1}(z)e^{-i(n+1)\phi _{0}}-J_{n-1}(z)e^{-i(n-1)\phi _{0}}\Big)\Big]\\ \nonumber &\times & (\slashed p_{+}+m_{e}) \Big[ \lambda_{0}^{\rho}\,J^{*}_{n}(z)e^{+in\phi _{0}}(z)+\lambda_{1}^{\rho}\,\,\frac{1}{2}\Big(J^{*}_{n+1}(z)e^{+i(n+1)\phi _{0}} + J^{*}_{n-1}(z)e^{+i(n-1)\phi _{0}}\Big)- \lambda_{2}^{\rho}\,\frac{1}{2\, i}\Big(J^{*}_{n+1}(z)e^{+i(n+1)\phi _{0}}\\ \nonumber &- & J^{*}_{n-1}(z)e^{+i(n-1)\phi _{0}}\Big)\Big]\Bigg]\times Tr\Bigg[ (\slashed k_{-}-m_{\mu})\gamma^{\nu} (\slashed k_{+}+m_{\mu})\gamma^{\beta}(g_{v}-g_{a}\gamma^{5}) \Bigg]
+\dfrac{e^{2}}{(q_{-}+q_{-}+nk)^{2}}\left(\dfrac{g^{2}}{16 \cos^{2}\theta_{w}}\right)\\ \nonumber &\times & \frac{1}{(q_{-}+q_{+}+nk)^{2}-M_{Z}^{2}}\Big(-g_{\mu\nu}+(1-\xi)\dfrac{(q_{-}+q_{+}+nk)_{\mu}(q_{-}+q_{+}+nk)_{\nu}}{M_{Z}^{2}}\Big)\\ \nonumber &\times &\Big(-g_{\rho\beta}+(1-\xi)\dfrac{(q_{-}+q_{+}+nk)_{\rho}(q_{-}+q_{+}+nk)_{\beta}}{(q_{-}+q_{+}+nk)^{2}}\Big) Tr\Bigg[(\slashed p_{-}-m_{e}) \Big[ \lambda_{0}^{\mu}\,J_{n}(z)e^{-in\phi _{0}}(z)\\ \nonumber &+ &\lambda_{1}^{\mu}\,\,\frac{1}{2}\Big(J_{n+1}(z)e^{-i(n+1)\phi _{0}} + J_{n-1}(z)e^{-i(n-1)\phi _{0}}\Big)+\lambda_{2}^{\mu}\,\frac{1}{2\, i}\Big(J_{n+1}(z)e^{-i(n+1)\phi _{0}}- J_{n-1}(z)e^{-i(n-1)\phi _{0}}\Big)\Big]\\ \nonumber &\times & (\slashed p_{+}+m_{e}) \Big[ \chi_{0}^{\rho}\,J^{*}_{n}(z)e^{+in\phi _{0}}(z)+\chi_{1}^{\rho}\,\,\frac{1}{2}\Big(J^{*}_{n+1}(z)e^{+i(n+1)\phi _{0}} + J^{*}_{n-1}(z)e^{+i(n-1)\phi _{0}}\Big) \\ \nonumber &- &\chi_{2}^{\rho}\,\frac{1}{2\, i}\Big(J^{*}_{n+1}(z)e^{+i(n+1)\phi _{0}}-J^{*}_{n-1}(z)e^{+i(n-1)\phi _{0}}\Big)\Big]\Bigg]\times Tr\Bigg[ (\slashed k_{-}-m_{\mu})\gamma^{\nu}(g_{v}-g_{a}\gamma^{5}) (\slashed k_{+}+m_{\mu})\gamma^{\beta} \Bigg]\Bigg\rbrace.
\end{eqnarray}
\normalsize
In equation (\ref{18}), the integral over $d^{3}k_{+}$ can be extended from three to four dimensions\cite{Greiner} such that:
\begin{equation}
\dfrac{d^{3}k_{+}}{ E_{\mu^{+}}}=2\,d^{4}k_{+}\delta^{4}(k_{+}^{2}-m_{\mu}^{2})\Theta((k_{+})_{0}).
\label{19}
\end{equation}
Thus, the integral part of the differential cross section becomes as follows:
\scriptsize
\begin{eqnarray}
\int_{}\dfrac{|\mathbf{k}_{-}|^{2}d|\mathbf{k}_{-}|d\Omega}{(2\pi)^2E_{\mu^{-}}}\int_{}\dfrac{d^{3}k_{+}}{ E_{\mu^{+}}}  \delta^{4}(k_{-}+k_{+}-q_{-}-q_{+}-nk)&=& \int_{}\dfrac{2\,|\mathbf{k}_{-}|^{2}d|\mathbf{k}_{-}|d\Omega}{(2\pi)^2E_{\mu^{-}}}\delta\Big((q_{+}+q_{-}+nk -k_{-})^{2}-m_{\mu}^{2}\Big)
\label{20}
\end{eqnarray}
\normalsize
The remaining integral over $d|\mathbf{k}_{-}|$ can be performed by using the well known formula \cite{Greiner} given by:
\begin{equation}
 \bigintsss d\mathbf y f(\mathbf y) \delta(g(\mathbf y))=\dfrac{f(\mathbf y)}{|g^{'}(\mathbf y)|_{g(\mathbf y)=0}}.
 \label{21}
\end{equation}
Therefore, the final expression of the differential cross section becomes as follows:
\small
\begin{eqnarray}
\dfrac{d\sigma_{n}}{d\Omega}({e}^{+}{e}^{-}\rightarrow \mu^{+}\mu^{-})&=&\dfrac{1}{16\sqrt{(q_{-}q_{+})^2-m_{e}^{*^{4}}}}   \big|\overline{A_{\gamma}^{n} + A_{Z}^{n}} \big|^{2} \dfrac{2|\mathbf{k}_{-}|^{2}}{(2\pi)^2E_{\mu^{-}}}\dfrac{1}{|g^{'}(|\mathbf{k}_{-}|)|_{g(|\mathbf{k}_{-}|)=0}},
\label{22}
\end{eqnarray}
\normalsize
where the expression of $g^{'}(|\mathbf{k}_{-}|)$ is given by:
\begin{equation}
\big|g^{'}(|\textbf{k}_{-}|)\big|=-2\Bigg[\Big[\sqrt{s}+n\omega-\frac{e^{2}a^{2}}{2}\Big(\dfrac{4}{\sqrt{s}}\Big)\Big]\dfrac{|\textbf k_{-}|}{\sqrt{|\textbf k_{-}|^{2}+m_{\mu}^{2}}}\Bigg].
\label{23}
\end{equation}
The total cross section is obtained by numerically integrating the equation (\ref{22}) over the solid angle $d\Omega$, and the trace caculation is performed by using Feyncalc Program \cite{Feyncalc}.
It is recognized that any computed observable, such as the cross section, does not depend on the parameter $\xi$ that appears in the propagators' expressions (equations (\ref{9}) and (\ref{10})). Thus, we have checked, numerically the gauge invariance of the laser-assisted total cross section, as it is illustrated in figure \ref{fig3}.
\section{Result and discussion}
In this paper's part, we have analyzed, in the centre of mass frame, the process which acts as a source of muon-antimuon pair production via electron-positron annihilation. 
The total cross section is obtained by numerically integrating the differential cross section, given by equation (\ref{22}), over the solid angle $d\Omega$. This cross section is analyzed as a function of the electromagnetic field parameters at different center of mass energies including the energy corresponding to the $Z$-boson peak. As we have mentioned above in Section (2), the laser beam is considered to be propagating along the $z$-axis while the colliding beam is propagating in the plan (oxy). Since the cross section is gauge invariant, and for simplicity reason, we have chosen the Feynman gauge ($\xi=1$).
The standard model parameters are taken from PDG \cite{PDG} such that: $m_{e} = 0.511\, MeV$, $m_{\mu} = 105.66\, MeV$, the mixing angle $\sin^{2}(\theta_{w})=0.23126$, the Fermi coupling constant $G_{F}=1.166 3787 \times 10^{-5}\, GeV^{-2}$, the $Z$-boson mass $M_{Z}=91.186\,GeV$ and its total width $\Gamma_{Z}=2.4952\,GeV$.
Before going any further in our discussion, we firstly want to check the validity of our results. To do this, we have compared the obtained laser-assisted total cross section with its corresponding laser-free total cross section. Thus, we have chosen the laser parameters as $n=0$ and  $\varepsilon_{0}=0\,\,V.cm^{-1}$.
\begin{figure}[H]
  \centering
      \includegraphics[scale=0.6]{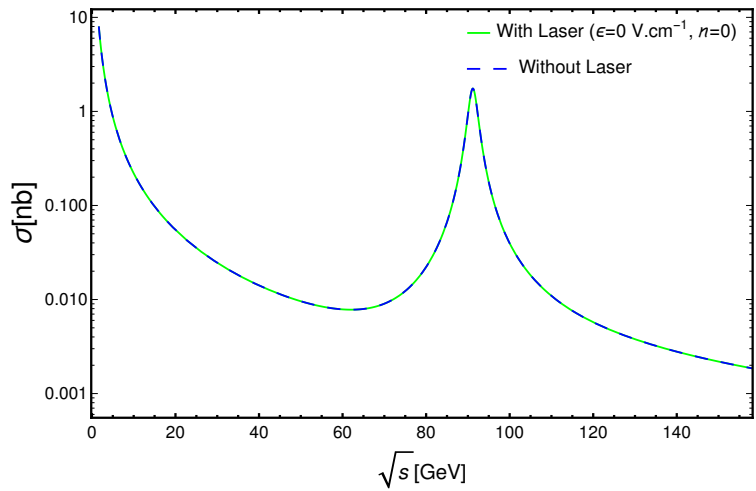}
  \caption{Comparison between the laser-free total cross section of muon pair production process with its corresponding laser-assisted total cross section by taking the laser parameters as: $\varepsilon_{0}=0\,V.cm^{-1} $ and $n=0$.}
  \label{fig4}
\end{figure}
Figure \ref{fig4} represents the comparison between the laser-free total cross section \cite{eeZH, Greiner} with its corresponding total cross section in the presence of a circularly polarized electromagnetic field. The fact that the analytical calculations, in the presence of a laser field, are too long makes this comparison as one of the most used techniques to check the validity of our calculations. According to this figure, it is clear that the two total cross sections are very consistent as they lead to the same results for every centre of mass energy. Therefore, this comparison confirms the validity of our results. The next step in this discussion concerns the behavior of the partial total cross section, which corresponds to each four-momentum conservation $p_{3}+p_{4}-q_{1}-q_{2}-nk=0$.
\begin{table}[H]
  \centering
\caption{\label{tab1}Laser-assisted partial total cross section versus of the number of emitted and absorbed photons for different laser field strengths and frequencies at $\sqrt{s}= M_{Z}$ .}
\begin{tabular}{cccccccc}
\hline
     & Total cross section[nb] & & Total cross section[nb] \\
   & $\varepsilon_{0}=10^{6}V.cm^{-1} $& &  $\varepsilon_{0}=10^{7}V.cm^{-1} $& \\
 n &$\omega=1.17eV $ & n & $\omega=1.17eV  $\\
 \hline
 $-150$ & $ 0 $  & $-1400$ & $ 0 $\\
  $-120$ & $ 0 $  & $-1100$ & $ 0 $\\
   $-110$ & $ 0 $ & $-800$ & $ 1.39631\times10^{-3} $\\  
    $-90$ & $ 2.29468\times10^{-2} $  & $-600$ & $ 1.29312\times10^{-3} $\\
     $-60$ & $ 1.29256\times10^{-2} $  & $-400$ & $ 3.67873\times10^{-4} $ \\
      $-30$ & $ 3.70605\times10^{-3} $ & $-200$ & $ 3.49014\times10^{-4} $\\
     $0$ & $ 4.42124\times10^{-3} $  & $0$ & $ 105399\times10^{-3} $\\
     $30$ & $ 3.70605\times10^{-3} $  & $200$ & $ 3.49014\times10^{-4} $\\
    $60$ & $ 1.29256\times10^{-2} $ & $400$ & $  3.67873\times10^{-4} $\\
     $90$ & $ 2.29468\times10^{-2} $ & $600$ & $ 1.29312\times10^{-3} $\\
   $110$ & $ 0 $   & $800$ & $ 1.39631\times10^{-3} $ \\
    $120$ & $ 0 $  & $1100$ & $ 0 $\\
    $150$ & $ 0 $  & $1400$ & $ 0 $\\
     \hline
\end{tabular}
\end{table}
Table \ref{tab1} illustrates the variation of the partial total cross section of the process $e^{+}e^{-}\rightarrow \mu^{+}\mu^{-}$ as a function of the number of exchanged photons for a given frequency, $\omega=1.17\,eV$, and for two typical laser field amplitudes which are $\varepsilon_{0}=10^{6}\,V.cm^{-1}$ and $\varepsilon_{0}=10^{7}\,V.cm^{-1}$. As we can see from this table, regardless of the laser field amplitude, the contribution of multiphoton processes presents two symmetric cutoffs with respect to $n=0$. These symmetric aspect is due to the presence of ordinary Bessel functions. In addition, the cross section which corresponds to negative value of $-n$ (emission of photons) is absolutely the same as that for $+n$ (absorption of photons). These cutoffs are successively $n=\pm 110$ and $n=\pm 1100$ for  $\varepsilon_{0}=10^{6}\,V.cm^{-1}$ and $\varepsilon_{0}=10^{7}\,V.cm^{-1}$. Above these cutoffs no photons will be exchanged between the electromagnetic field and the colliding particles. We remark also that the cutoff value increases as far as we increase the strength of the laser field. To understand more clearly this behavior, we have plotted in figure \ref{fig5}, the partial total cross section versus the laser photon's number for $\varepsilon_{0}=10^{6}\,V.cm^{-1}$ and for two different known laser frequencies.
\begin{figure}[t]
  \centering
      \includegraphics[scale=0.6]{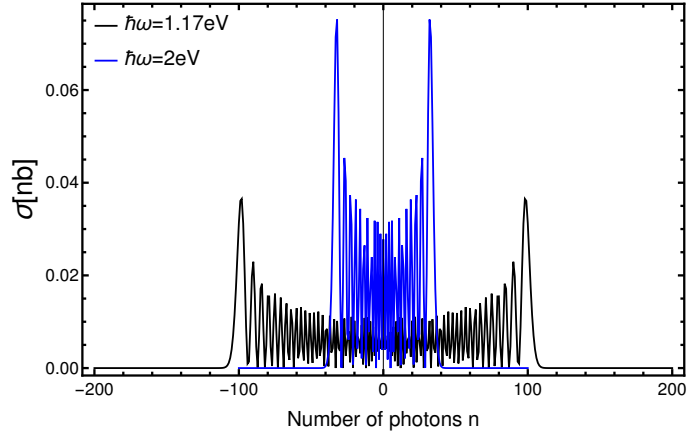}
  \caption{Dependence of the laser-assisted partial total cross section on the number of exchanged photons for two laser frequencies, and by taking the centre of mass energy as $\sqrt{s}=M_{Z}$. The laser field amplitude is chosen as $\varepsilon_{0}=10^{6}\,V.cm^{-1}$.}
  \label{fig5}
\end{figure}
Curves presented in figure \ref{fig5} confirms that obtained in table \ref{tab1} as the cutoffs number for $\varepsilon_{0}=10^{6}\,V.cm^{-1}$ and $\omega=1.17\,eV$ are approximately $n=\pm110$. However for $\varepsilon_{0}=10^{6}\,V.cm^{-1}$ and $\omega=2\,eV$, the cutoffs number are approximately $n=\pm50$. Therefore, the number of photons, which is required to be transferred between the laser beam and the colliding system, decreases by increasing the frequency of the laser radiation. An other important point to be mentioned here is that the order of magnitude of the cross section increases as a function of the laser frequency. Let's focus, now, on the behavior of the total cross section, which is summed over a number of laser photons, for a wide range of centre of mass energies.
\begin{figure}[H]
  \centering
      \includegraphics[scale=0.50]{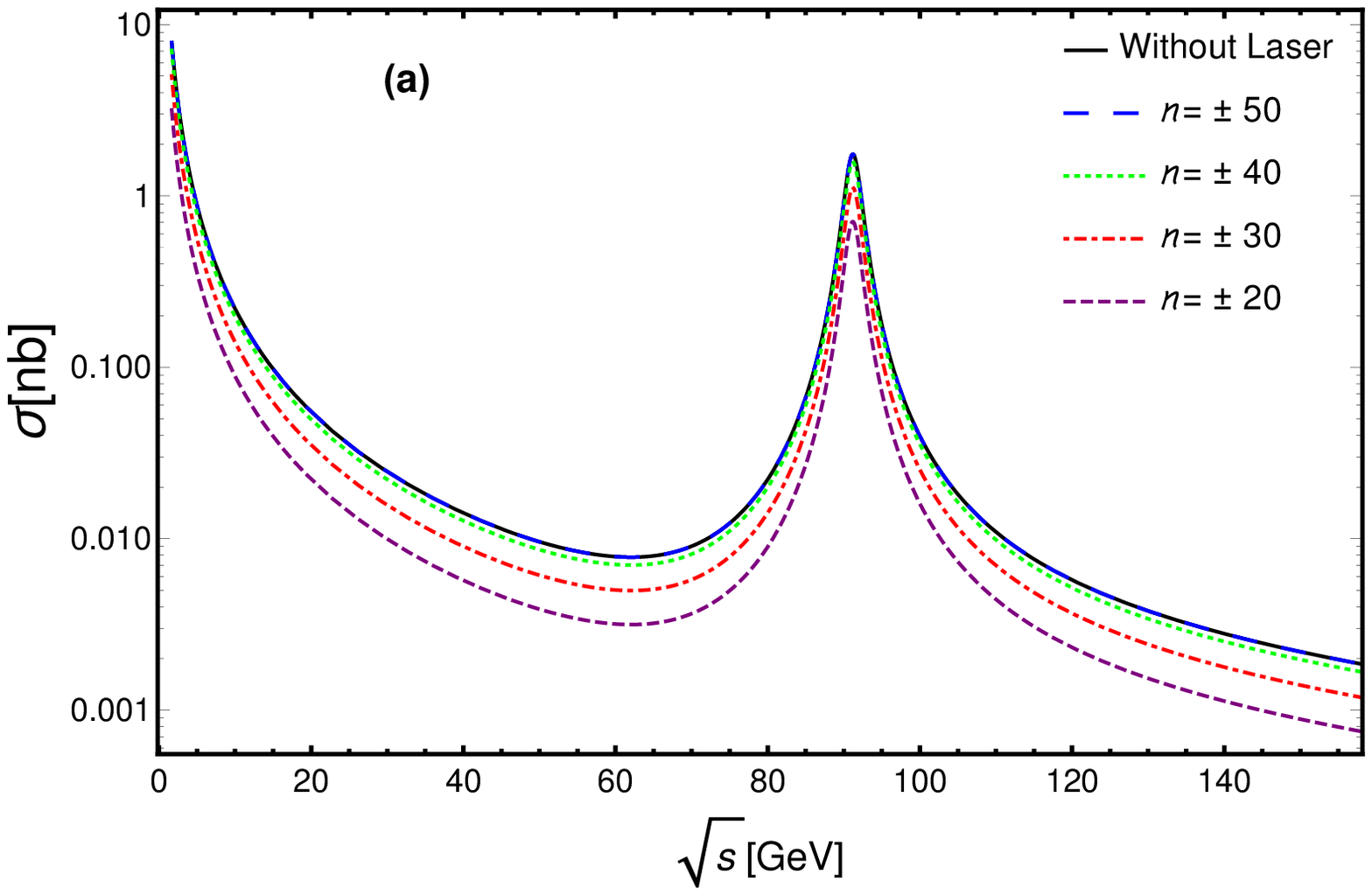}\hspace*{0.4cm}
      \includegraphics[scale=0.50]{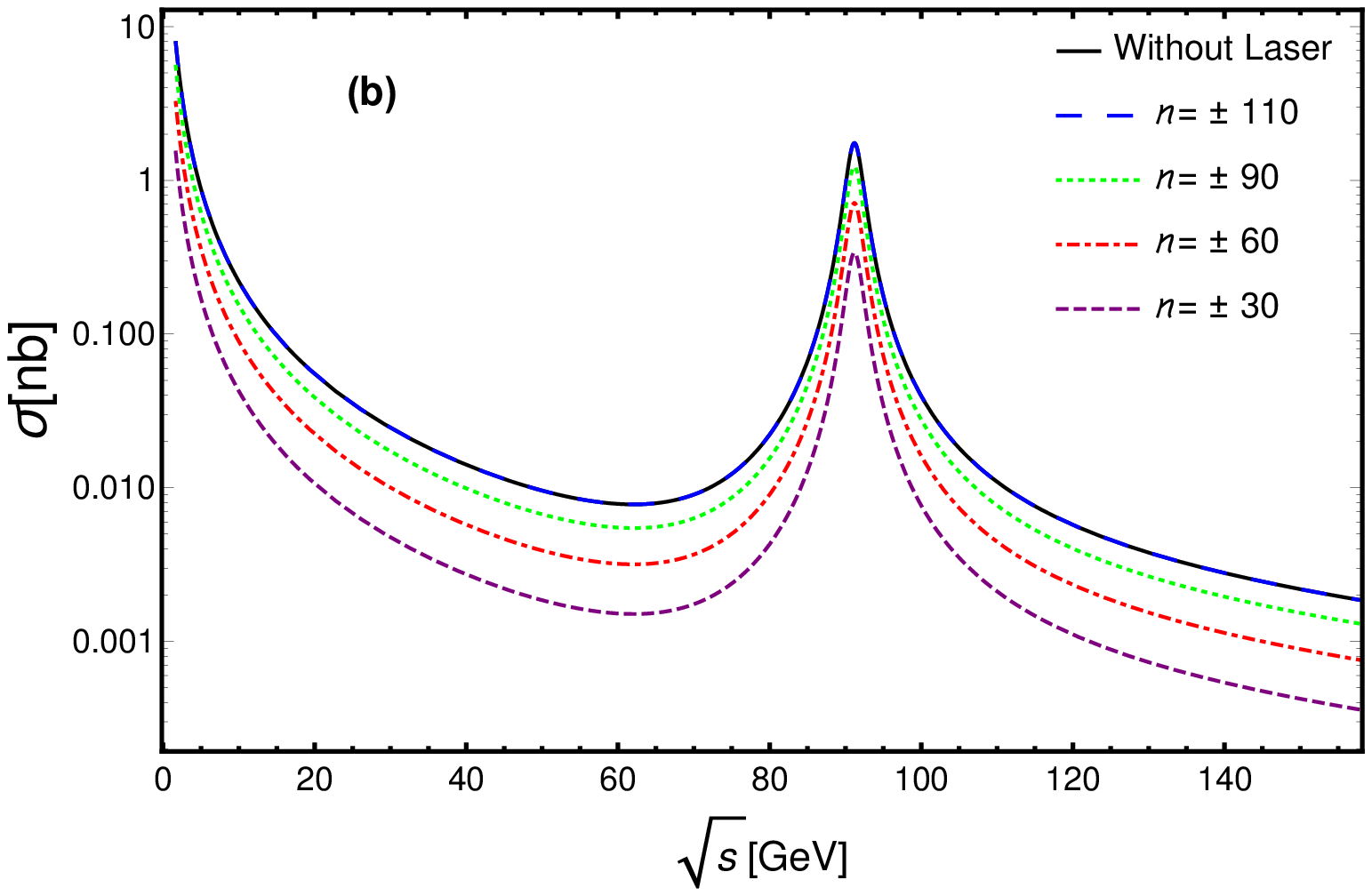}\par\vspace*{0.5cm}
        \caption{Dependence of the $e^{+}e^{-}\rightarrow \mu^{+}\mu^{-}$ total cross section on the centre of mass energy for different number of exchanged photons by choosing the laser field strength as $\varepsilon_{0}=10^{6}\,V.cm^{-1}$ and its frequency as $\omega=2\,eV$ in (a) and $\omega=1.17\,eV$ in (b).}
        \label{fig6}
\end{figure} 
Figure \ref{fig6} displays the variation of the total cross section as a function of the collider centre of mass energy for different transferred photons number and for two different laser frequencies. In the absence of an external field,
the QED process dominates at low centre of mass energies, $\sqrt{s}< M_{Z}$, due to the presence of the $M_{Z}^{2}$ term in the $Z$-boson propagator. However, at very high centre of mass energies where $\sqrt{s}> M_{Z}$, the QED and $Z$-boson exchange processes are both important because the strengths of the couplings of the photon and the $Z$-boson are comparable. In the region $\sqrt{s}= M_{Z}$, the $Z$-boson process dominates. In the presence of the laser field, the general aspect of the total cross section doesn't change for all centre of mass energies. However, we remark that it decreases by several orders of magnitude as compared to its corresponding laser-free cross section. Indeed, it raises as far as we increase the number of photons exchanged. In addition, the summation over $\pm$ cutoff, as $n=\pm 50$ in (a) or $n=\pm 110$ in (b), leads to a cross section which is equal to the laser-free cross section in all centre of mass energies. This result is called sum-rule, and it is elaborated by Kroll and Watson in \cite{Watson}. By comparing figure \ref{fig6}(a) and figure \ref{fig6}(b), we observe that the required number of photons to be transferred in order to reach the sum-rule increases by decreasing the laser field frequency.
For instance, at $\sqrt{s}=60\,GeV$ and for $n=\pm\,30$, the total cross-section is equal to $0.00506884\,[nb]$ and $0.00160037\,[nb]$ for the laser frequencies $\omega=2\,eV$ (Fig. \ref{fig6}(a)) and $\omega=1.17\,eV$ (Fig. \ref{fig6}(b)), respectively. 
No, let's focus our attention on how the total cross section behaves as a function of the centre of mass energy for different laser field strengths and for different frequencies.
\begin{figure}[H]
  \centering
      \includegraphics[scale=0.50]{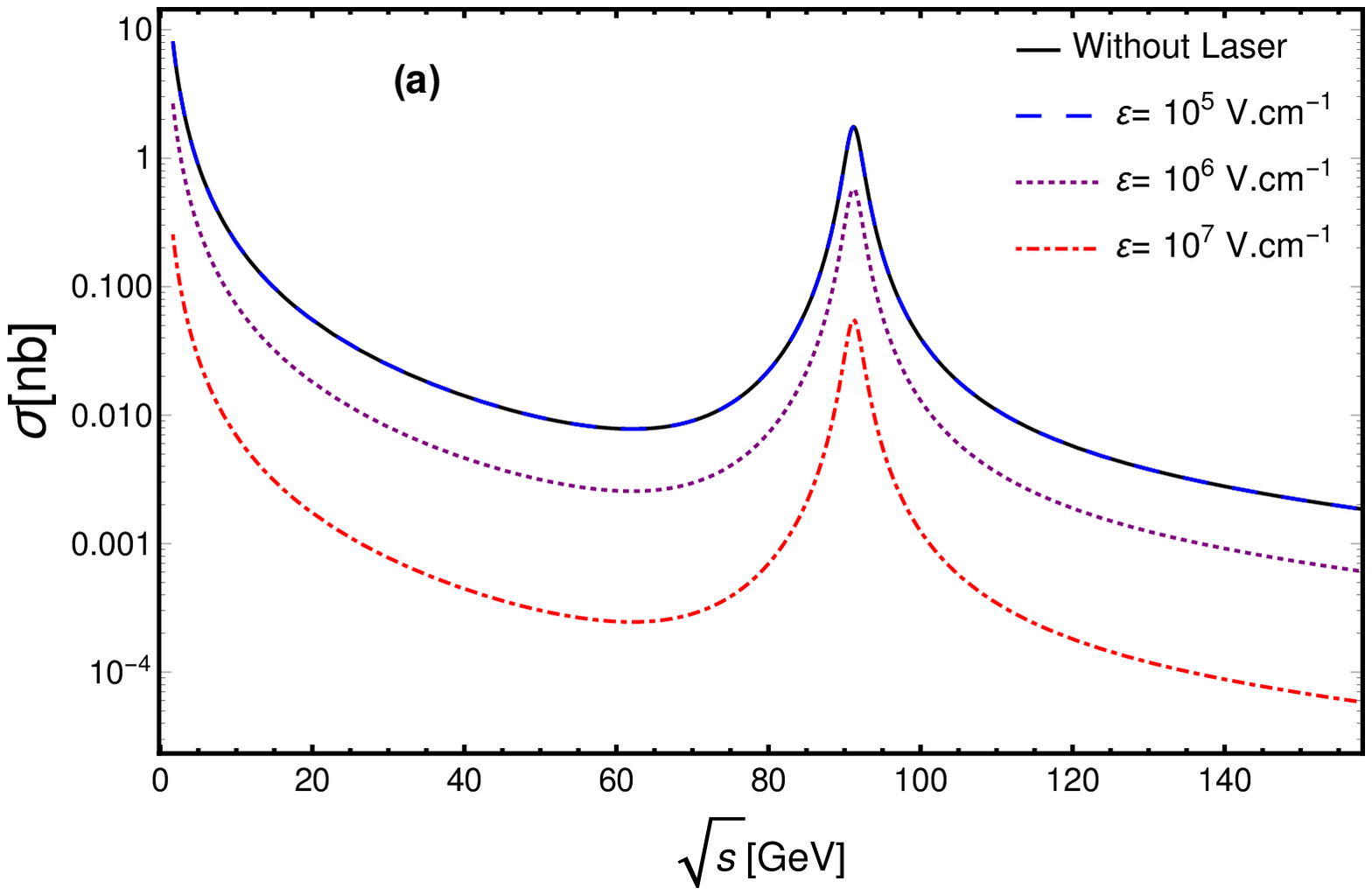}\hspace*{0.4cm}
      \includegraphics[scale=0.50]{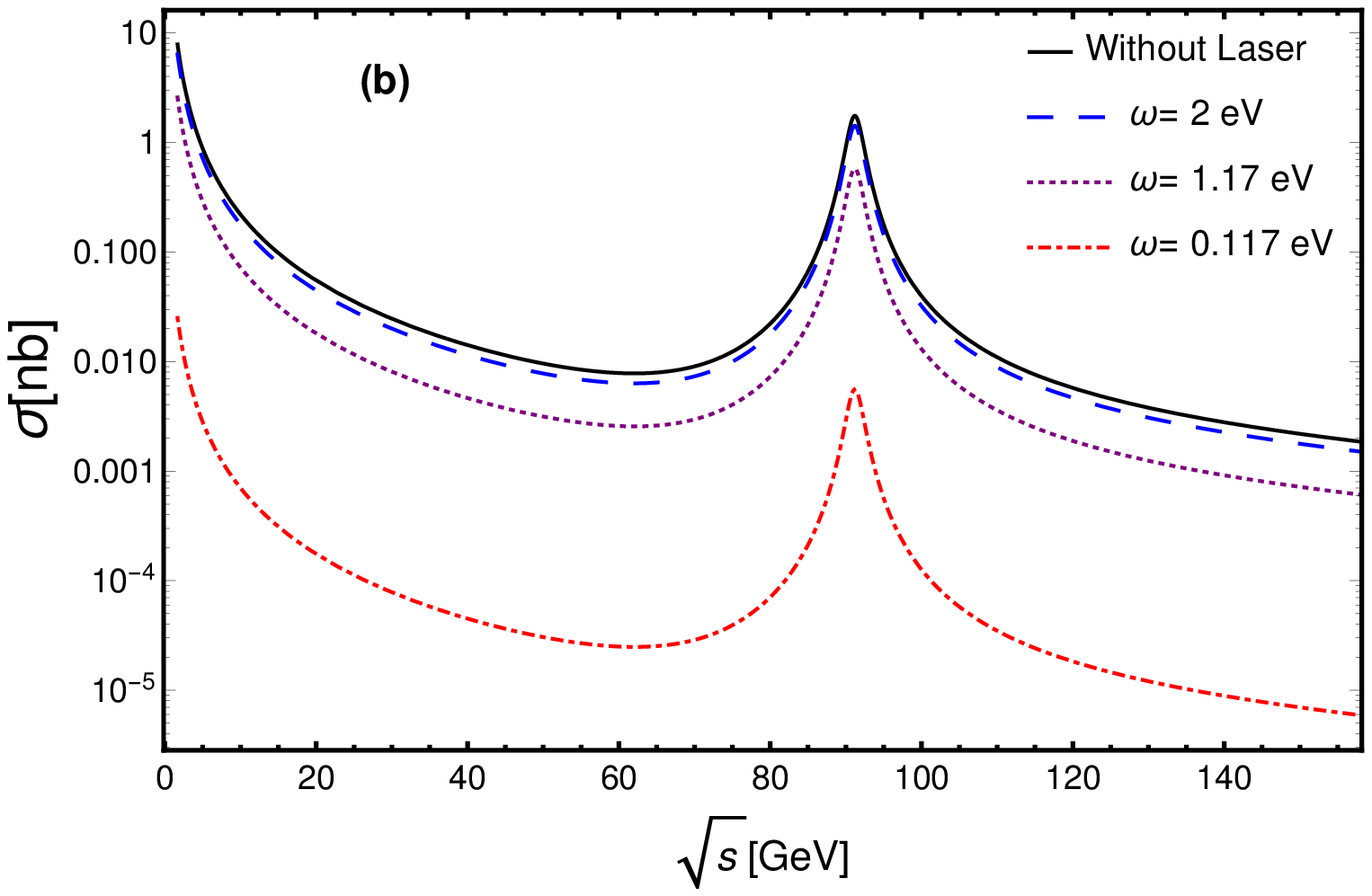}\par\vspace*{0.5cm}
        \caption{Variation of the $e^{+}e^{-}\rightarrow \mu^{+}\mu^{-}$ total cross section as a function of the centre of mass energy by taking $n=\pm\,40$. (a): $\omega=1.17\,eV$ and different laser field amplitudes. (b): $\varepsilon=10^{6}\,V.cm^{-1}$  and different laser frequencies.}
        \label{fig7}
\end{figure}
Figure \ref{fig7} represents the variation of the total cross-section as a function of the centre of mass energy for different laser field strengths (Fig. \ref{fig7}(a)) and for different laser frequencies (Fig. \ref{fig7}(b)). To avoid intensive calculations, we have chosen $n$ as $\pm\,40$. According to (Fig. \ref{fig7}(a)), the general aspect of the total cross section remains the same. However, the order of magnitude of the cross section decreases as long as the laser field strength increases. For instance, at $\sqrt{s}=60\,GeV$, its value is equal to $0.0835682\,[nb]$, $0.00269587\,[nb]$ and $0.000259496\,[nb]$ for $\varepsilon_{0}=10^{5}\,V.cm^{-1}$, $\varepsilon_{0}=10^{6}\,V.cm^{-1}$  and $\varepsilon_{0}=10^{7}\,V.cm^{-1}$, respectively. From (Fig. \ref{fig7}(b)), we conclude that the total cross section decreases by decreasing the laser field frequency. For example, at $\sqrt{s}=60\,GeV$, the total cross section value is successively equal to $0.00641201\,[nb]$, $0.002758\,[nb]$ and $0.0000254119\,[nb]$ for the frequencies $2\,eV$, $1.17\,eV$ and $0.117\,eV$. The effect of the laser field amplitude on the cross section is illustrated more clearly in figure \ref{fig8}.
\begin{figure}[t]
  \centering
      \includegraphics[scale=0.6]{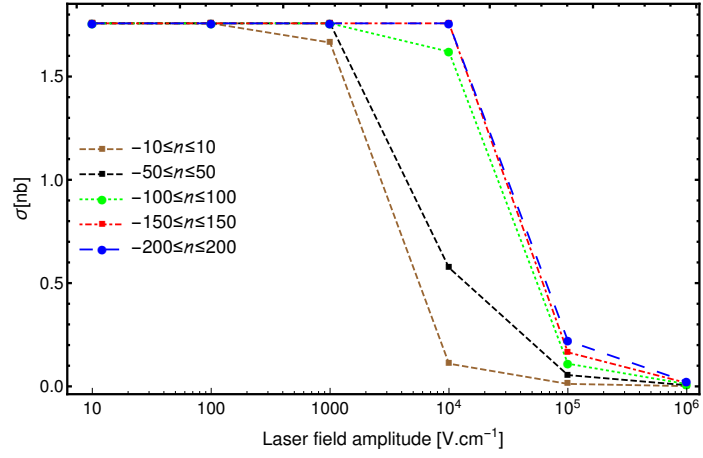}
  \caption{Variation of the laser-assisted total cross section as a function of the laser field amplitude at $\sqrt{s}=M_{Z}$ for different number of exchanged photons and by taking the laser frequency as $\omega=0.117\,eV$.}
  \label{fig8}
\end{figure}
Figure \ref{fig8} displays the behavior of the total cross section versus the laser field strength for different number of exchanged photons at the $Z$-boson peak. In the region of low electromagnetic amplitude, we observe that the laser field doesn't affect the cross section regardless of the transferred number of photons. In addition, this region extends as much the laser photon's number increases. Therefore, as the laser amplitude reaches a threshold value, the total cross section begins to decrease. This decreasing process continues as long as we increase the laser field amplitude until it becomes zero. However, if we sum over $n$ from $-$cutoff to $+$cutoff, the cross section will not show any dependence on the electromagnetic field regardless of its strength. Let's move now to study the behavior of the total cross-section in a wide range of centre of mass energies including below and above the $Z$-boson peak.
\begin{figure}[H]
  \centering
      \includegraphics[scale=0.6]{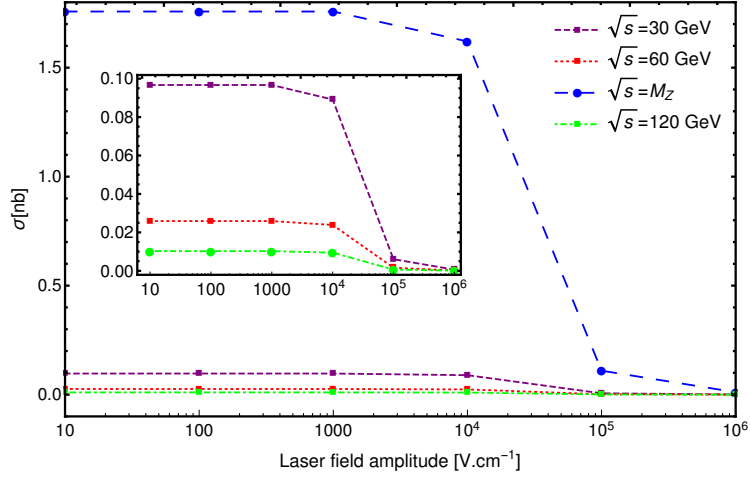}
  \caption{Dependence of the laser-assisted total cross section on the laser field amplitude for different centre of mass energies by choosing $n$ as $\pm\,100$ and $\omega$ as $0.117\,eV$.}
  \label{fig9}
\end{figure}
Figure \ref{fig9} represents the variation of the total cross-section as a function of the laser field strength for a range of centre of mass energies between $30\,GeV$ and $120\,GeV$ by taking the number of exchanged photons as $n=\pm\,100$ and the laser frequency as $\omega=0.117\,eV$. It is obvious that, for each centre of mass energy, there is a threshold value of the laser field amplitude from which the electromagnetic field begins to affect the cross section. Above this threshold value, the total cross-section decreases progressively until it becomes zero. This result is in full agreement with that found in figure \ref{fig8}.
An other important remark is that the order of magnitude of the total cross-section depends also on the centre of mass energy as it is high for the energy corresponding to the $Z$-boson peak, and decreases outside the peak.
To study the total cross-section simultaneous dependence on the laser field strength and its frequency, we have plotted its variation in a contour plot for two centre of mass energies which are $\sqrt{s}= M_{Z}$ and $\sqrt{s}=40\, GeV$.
\begin{figure}[H]
 \centering
     \includegraphics[scale=0.6]{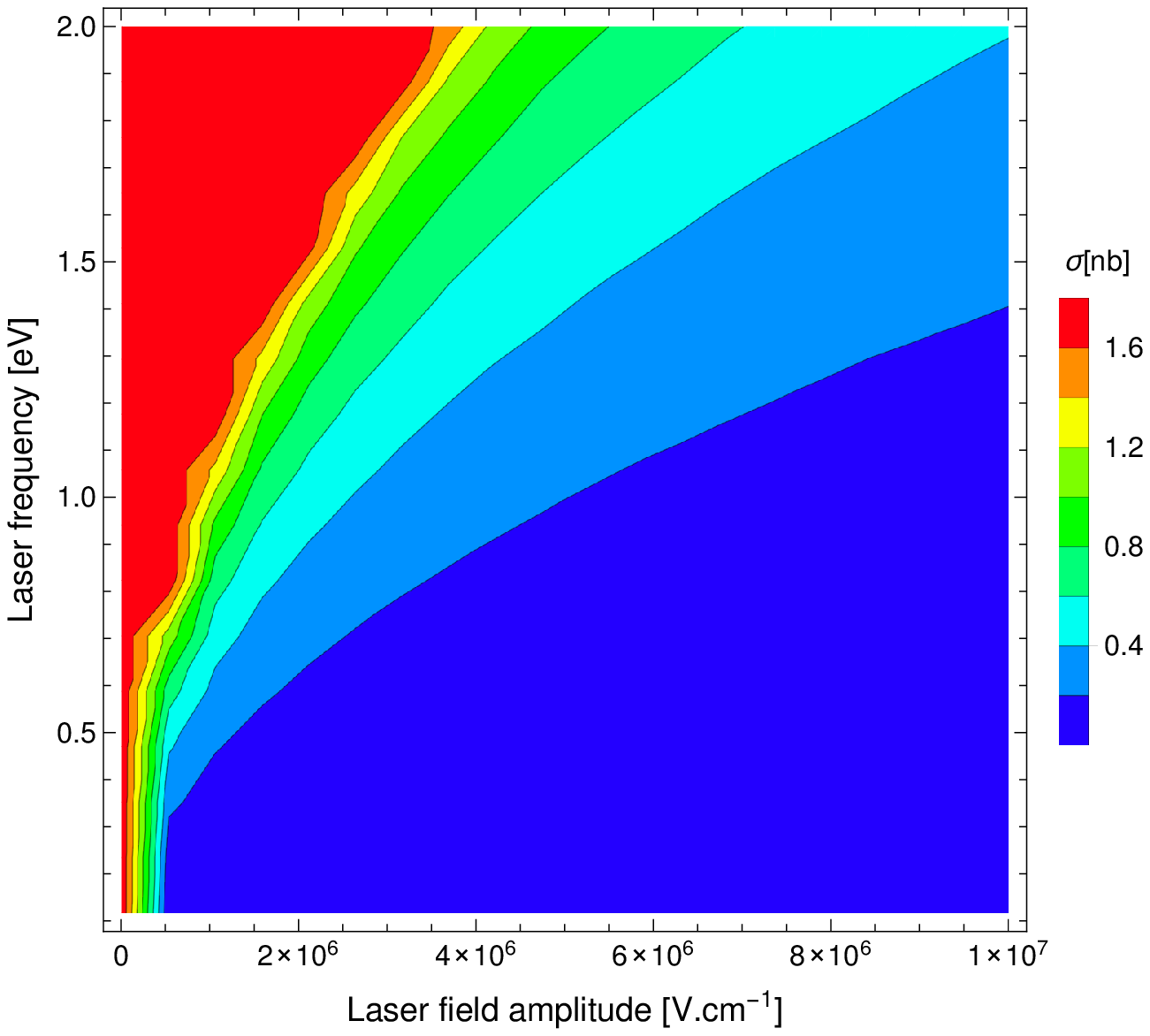}\hspace*{0.4cm}
     \includegraphics[scale=0.7]{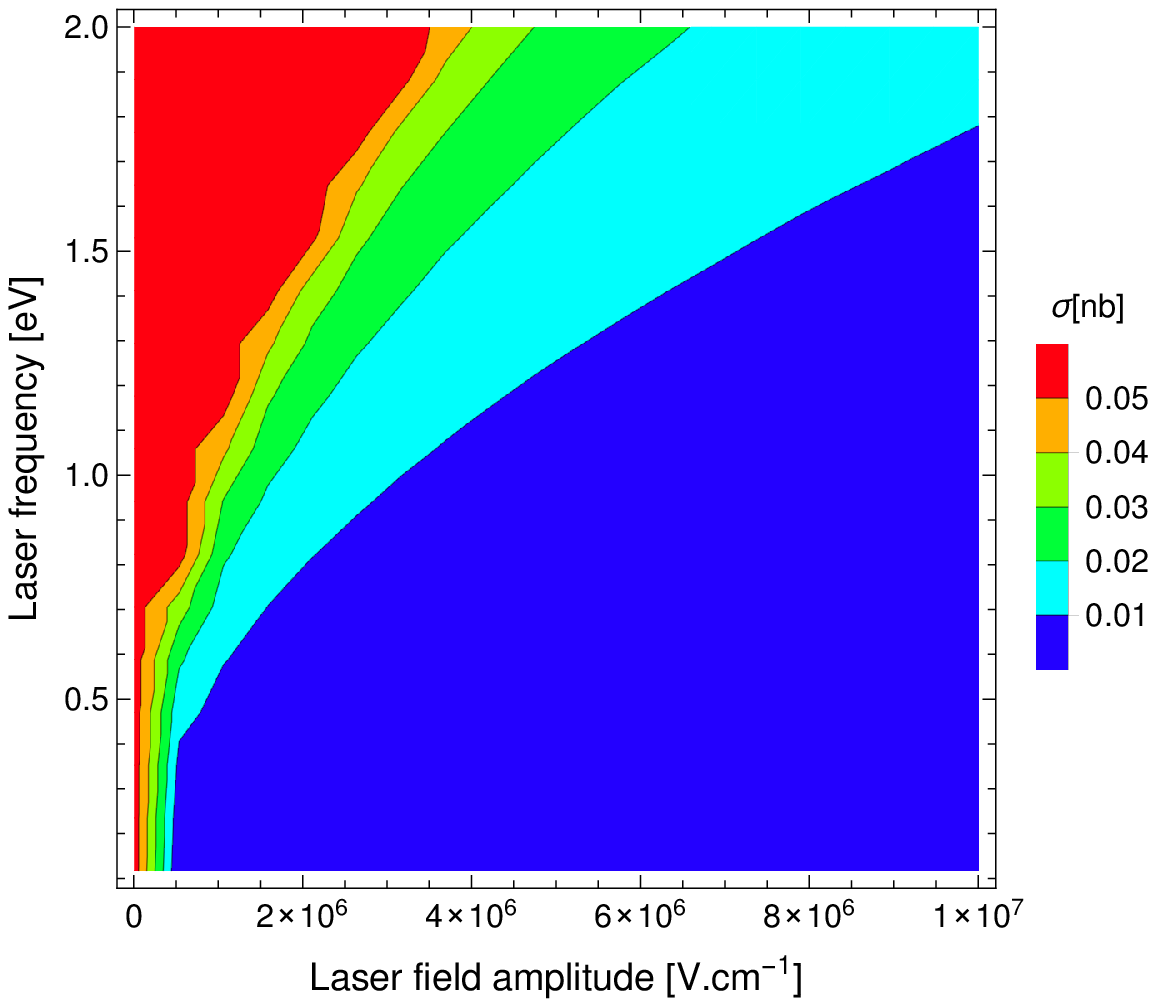}\par\vspace*{0.5cm}
        \caption{Behavior of the laser-assisted total cross section as a function of the laser field strength and its frequency by taking the number of transferred photons as $n=\pm\,120$ and the centre of mass energy as $\sqrt{s}= M_{Z}$ in the left panel and $\sqrt{s}=40\, GeV$ in the right panel.}
       \label{fig10}
\end{figure}
Figure \ref{fig10} displays some values of the total cross section for different combinations of the laser field amplitude and its frequency. To avoid intensive calculation, we have summed over the number of transferred photons from $-120$ to $+120$. These contour plots confirms the results obtained in figure \ref{fig6} for that the total cross section decreases as much as the laser strength increases or by decreasing the laser frequency. Indeed, the highest laser-assisted cross section occurs at low laser field strengths and high laser field frequencies.
\section{Conclusion}
In this paper, we have investigated the process of muon pair production in the presence of a circularly polarized laser field. The calculations of the total cross section is performed for the leading order by taking into account the interference between $\gamma$ and $Z$ diagrams. We have found that radiations with small intensities have no effect on the cross section. However, for high intensities, the laser field decreases the total cross section by several orders which depend on the number of exchanged photons, laser field strength and its frequency. We have also shown that the circularly polarized laser field strongly affects the production process at high laser field strengths and/or low frequencies. Therefore, the probability of muon-antimuon creation is reduced in the presence of an electromagnetic field with circular polarization.


\begin{thebibliography}{99}
\bibitem{1}P. Francken, C. J. Joachain, Electron — atomic-hydrogen elastic collisions in the presence of a laser field, Phys. Rev. A \textbf{35}, 1590 (1987). \url{https://doi.org/10.1103/PhysRevA.35.1590}; P. Francken, Y. Attaourti, and C. J. Joachain
Phys. Rev. A \textbf{38}, 1785(1988). \url{https://doi.org/10.1103/PhysRevA.38.1785}.  


\bibitem{2}Y. Attaourti and B. Manaut, Comment on “Mott scattering in strong laser fields", Phys. Rev. A \textbf{68}, 067401. \url{https://doi.org/10.1103/PhysRevA.68.067401};
Y. Attaourti, B. Manaut, and S. Taj, Mott scattering in an elliptically polarized laser field, Phys. Rev. A \textbf{70}, 023404. \url{https://doi.org/10.1103/PhysRevA.70.023404};
Y. Attaourti, B. Manaut, and A. Makhoute, Relativistic electronic dressing in laser-assisted electron-hydrogen elastic collisions, Phys. Rev. A \textbf{69}, 063407. \url{https://doi.org/10.1103/PhysRevA.69.063407};
Y. Attaourti, S. Taj, and B. Manaut, Semirelativistic model for ionization of atomic hydrogen by electron impact, Phys. Rev. A \textbf{71}, 062705. \url{https://doi.org/10.1103/PhysRevA.71.062705};
B. Manaut, S. Taj, and Y. Attaourti, Mott scattering of polarized electrons in a strong laser field, Phys. Rev. A \textbf{71}, 043401 . \url{https://doi.org/10.1103/PhysRevA.71.043401}.

\bibitem{3}E. Hrour, M. El Idrissi, S. Taj and B. Manaut, Relativistic elastic scattering of hydrogen atoms by positron impact with anomalous magnetic moment effects, Ind. J. Phys \textbf{89}, 783-788 (2015). \url{https://doi.org/10.1007/s12648-015-0649-0};
B Manaut, Y Attaourti, S Taj and S. Elhandi, Mott scattering of polarized electrons in a circularly polarized laser field, Phys. Scr. \textbf{80}, 2009  025304. \url{https://doi.org/10.1088/0031-8949/80/02/025304};
E. Hrour, S. Taj, A. Chahboune and B. Manaut, "Relativistic proton-impact excitation of hydrogen atom in the presence of intense laser field." Can. J. Phys, \textbf{94}, 7 (2016);
S. Taja, B. Manaut, M. El Idrissi, Y. Attaourti, and L. Oufni, J. At. Mol. Sci \textbf{4}, 18-29 (2013). \url{https://doi: 10.4208/jams.030112.032212a}.


\bibitem{4} G. A. Mourou, T. Tajima, and S. V. Bulanov, Rev. Mod.Phys. \textbf{78}, (2006) 309-371. \url{https://doi.org/10.1103/RevModPhys.78.309}.

\bibitem{5}M. Ouhammou, M. Ouali, S. Taj, and B. Manaut, Laser Phys. Lett. \textbf{18}, 076002 (2021). \url{https://doi.org/10.1088/1612-202X/ac0919}.

\bibitem{6} Sarah J. Müller, Christoph H. Keitel, Carsten Müller, Higgs boson creation in laser-boosted lepton collisions, Phys. Lett. B \textbf{730} (2014) 161–165. \url{https://doi.org/10.1016/j.physletb.2014.01.047}.

\bibitem{7} M. Ouali, M. Ouhammou, S. Taj, R. Benbrik, B. Manaut, Laser-assisted charged Higgs pair production in Inert Higgs Doublet Model (IHDM). \url{https://arxiv.org/abs/2109.05530}.

\bibitem{8} M. Ouhammou, M. Ouali, S. Taj, B. Manaut, 2021, Chin. J. Phys, \url{https://doi.org/10.1016/j.cjph.2021.09.012}.

\bibitem{9} Muller, K. Z. Hatsagortsyan, and C. H. Keitel, Phys. Rev. D \textbf{74}, 074017 (2006). \url{https://doi.org/10.1103/PhysRevD.74.074017}.

\bibitem{10} Carsten Muller, Karen Z. Hatsagortsyan, and Christoph H. Keitel, 2008, Phys. Rev. A, \textbf{78}, 033408. \url{https://doi.org/10.1103/PhysRevA.78.033408}.

\bibitem{11} M. Ouali, M. Ouhammou, S. Taj, B. Manaut, $Z$-boson production via the weak process ${e}^{+} {e}^{-}\rightarrow {\mu}^{+} {\mu}^{-}$ in the presence of a circularly polarized laser field, arXiv:2105.14854 [hep-ph]. \url{https://arxiv.org/abs/2105.14854}.

\bibitem{12} Nikishov A~I~ and Ritus V~I~1964 \emph{Zh. Eksp. Teor. Fiz.} \textbf{46}, 776 and 1768 [1964 Sov. Phys.-JETP \textbf{19}, 529 and 1191]; Denisov~M~M~and Fedorov~M~V~1967 \emph{Zh. Eksp. Teor. Fiz.} \textbf{53}, 1340-1348; F Ehlotzky, Krajewska~K and Kami\'{n}ski~J~Z 2009 \emph{Rep. Prog. Phys.} \textbf{72} 046401.

\bibitem{13} M Baouahi \textit{et al.}, Laser-assisted kaon decay and CPT symmetry violation, Laser Phys. Lett. \textbf{18}, 106001 (2021). \url{https://doi.org/10.1088/1612-202X/ac1e86}.

\bibitem{14} Sarah J. Müller, Christoph H. Keitel, and Carsten Müller, Particle production reactions in laser-boosted lepton collisions, Phys. Rev. D \textbf{90}, 094008 (2014). \url{https://doi.org/10.1103/PhysRevD.90.094008}.

\bibitem{15} M.Jakha, S.Mouslih, S.Taj, Y.Attaourti, B.Manaut, Influence of intense laser fields on measurable quantities in $W^{-}$-boson decay. \url{https://doi.org/10.1016/j.cjph.2021.09.011}.

\bibitem{16} Sarah J. Müller, Electroweak Processes in Laser-Boosted Lepton Collisions, 2013, PhD thesis, Ruperto-Carola University, Heidelberg, Germany. \url{https://doi.org/10.1142/9789814689304_0009}.

\bibitem{17} Nikita R. Larin, Victor V. Dubov, and Sergei P. Roshchupkin, Resonant photoproduction of high-energy electron-positron pairs in the ﬁeld of a nucleus and a weak electromagnetic wave, Phys. Rev. A \textbf{100}, 052502 (2019). \url{https://doi.org/10.1103/PhysRevA.100.052502}.

\bibitem{ILC-CLC}
L.~Linssen, A.~Miyamoto, M.~Stanitzki and H.~Weerts,
\textit{`Physics and Detectors at CLIC: CLIC Conceptual Design Report,''}
doi:10.5170/CERN-2012-003,
[arXiv:1202.5940 [physics.ins-det]];
H.~Baer, T.~Barklow, K.~Fujii, Y.~Gao, A.~Hoang, S.~Kanemura, J.~List, H.~E.~Logan, A.~Nomerotski and M.~Perelstein, \textit{et al.}
\textit{``The International Linear Collider Technical Design Report - Volume 2: Physics,''}
[arXiv:1306.6352 [hep-ph]].

\bibitem{FCC-CEPC} J.~B.~Guimar\~aes da Costa \textit{et al.} [CEPC Study Group],
\textit{``CEPC Conceptual Design Report: Volume 2 - Physics \& Detector,''},
[arXiv:1811.10545 [hep-ex]];
 [CEPC Study Group],
\textit{``CEPC Conceptual Design Report: Volume 1 - Accelerator,''},
[arXiv:1809.00285 [physics.acc-ph]]

\bibitem{circ} Muller, K. Z. Hatsagortsyan, and C. H. Keitel, Phys. Rev. D \textbf{74}, 074017 (2006). \url{https://doi.org/10.1103/PhysRevD.74.074017}.

\bibitem{linear}Carsten Muller, Karen Z. Hatsagortsyan, and Christoph H. Keitel, Muon pair creation from positronium in a linearly polarized laser field, 2008, Phys. Rev. A, \textbf{78}, 033408. \url{https://doi.org/10.1103/PhysRevA.78.033408}.

\bibitem{Volkov} D. M. Volkov, Uber eine Klasse von Losungen der Diracschen Gleichung, Z. Phys. \textbf{94}, (1935) 250-260. \url{http://dx.doi.org/10.1007/BF01331022}.

\bibitem{Greiner}W. Greiner and B. Mueller, Gauge Theory of Weak Interactions, 3rd ed. (Springer, Berlin, 2000).

\bibitem{Auger}F.V. Bunkin, M.V. Fedorov, Zh. Eksp. Teor. Fiz. \textbf{49}, 1215 (1965).

\bibitem{Feyncalc} R. Mertig, M. Bohm, and A. Denner, FEYN CALC: Computer algebraic calculation of feynman amplitudes, Comput. Phys. Commun. \textbf{64}, (1991) 345-359. \url{http://dx.doi.org/10.1016/0010-4655(91)90130-D}; V. Shtabovenko, R. Mertig, and F. Orellana, New developments in FeynCalc 9.0, Comput. Phys. Commun. \textbf{207}, (2016) 432-444. \url{http://dx.doi.org/10.1016/j.cpc.2016.06.008}.

\bibitem{PDG} P.A. Zyla et al. (Particle Data Group), Review of Particle Physics, 2020, Prog. Theor. Exp. Phys. \textbf{2020}, 083C01. \url{https://doi.org/10.1093/ptep/ptaa104}.

\bibitem{eeZH} Xin Mo et al, Chin. Phys. C \textbf{40}, 033001 (2016). \url{https://doi.org/10.1088/1674-1137/40/3/033001}.

\bibitem{Watson}F. V. Bunkin and M. V. Fedorov, Bremsstrahlung in a Strong Radiation Field, Sov. Phys. JETP \textbf{22}, (1966) 844-847; N. M. Kroll and K. M. Watson, Charged-Particle Scattering in the Presence of a Strong Electromagnetic Wave, Phys. Rev. A \textbf{8}, (1973) 804-809. \url{https://doi.org/10.1103/PhysRevA.8.804}.

\end{thebibliography}
\end{document}